\documentclass[twocolumn,aps,prc,superscriptaddress,showpacs,floatfix,nofootinbib]{revtex4-1}
\usepackage{url}
\usepackage{cancel}
\usepackage[colorlinks,linkcolor=blue,citecolor=blue,filecolor=black,urlcolor=blue]{hyperref}
\usepackage{epsfig,graphics}
\usepackage{graphicx}
\usepackage{dcolumn}
\usepackage{bm}
\usepackage[usenames]{color}
\usepackage{amssymb}
\usepackage{amsmath}
\usepackage{multirow}
\usepackage{float}
\usepackage{harpoon}
\usepackage{MnSymbol}
\usepackage{appendix}
\usepackage{color}
\usepackage{hyperref}
\usepackage{cleveref}

\newcommand{\sqrtsnn}{\mbox{$\sqrt{s^{}_{\mathrm{NN}}}$}}

\newcommand{\auau}{$^{197}$Au+$^{197}$Au}

\newcommand{\nch}{N_{\mathrm{ch}}}

\begin{document}
\title{Probing configuration of $\alpha$ clusters with spectator particles in relativistic heavy-ion collisions}
\author{Lu-Meng Liu}
\affiliation{Physics Department and Center for Particle Physics and Field Theory, Fudan University, Shanghai 200438, China}
\author{Song-Jie Li}\email[Song-Jie Li and Lu-Meng Liu contributed equally to this work.]{}
\affiliation{School of Physics Science and Engineering, Tongji University, Shanghai 200092, China}
\author{Zhen Wang}
\affiliation{School of Physics Science and Engineering, Tongji University, Shanghai 200092, China}
\author{Jun Xu}\email[Correspond to\ ]{junxu@tongji.edu.cn}
\affiliation{School of Physics Science and Engineering, Tongji University, Shanghai 200092, China}
\author{Zhong-Zhou Ren}\email[Correspond to\ ]{zren@tongji.edu.cn}
\affiliation{School of Physics Science and Engineering, Tongji University, Shanghai 200092, China}
\author{Xu-Guang Huang}
\affiliation{Physics Department and Center for Particle Physics and Field Theory, Fudan University, Shanghai 200438, China}
\affiliation{Key Laboratory of Nuclear Physics and Ion-beam Application (MOE), Fudan University, Shanghai 200433, China}
\affiliation{Shanghai Research Center for Theoretical Nuclear Physics,National Natural Science Foundation of China and Fudan University, Shanghai 200438, China}
\date{\today}
\begin{abstract}
We propose to use spectator particle yield ratios to probe the configuration of $\alpha$ clusters in $^{12}$C and $^{16}$O by their ultracentral collisions at RHIC and LHC energies. The idea is illustrated based on initial density distributions with various $\alpha$-cluster configurations generated by a microscopic cluster model, and without $\alpha$ clusters from mean-field calculations. The multifragmentation of the spectator matter produces more spectator light nuclei including $\alpha$ clusters in collisions of nuclei with chain structure of $\alpha$ clusters, compared to those of nuclei with a more compact structure. The yield ratio of free spectator neutrons to spectator particles with mass-to-charge ratio $A/Z=2$ scaled by their masses can be practically measured by the zero-degree calorimeter (ZDC) at RHIC and LHC, serving as a clean probe free from modeling the complicated dynamics at midrapidities.
\end{abstract}
\maketitle


Structure of nuclei is a fundamental problem in nuclear physics, and probing it by using relativistic heavy-ion collisions has become a hot topic in recent years. Basically, the density distribution of the colliding nuclei determines the initialization of the quark-gluon plasma (QGP) and thus the phase-space distribution of final produced particles. Deformation of colliding nuclei may have special features in the initial state of the collision and thus the final anisotropic flows~\cite{Filip:2009zz,Shou:2014eya,Giacalone:2019pca,Zhang:2021kxj}. The neutron-skin thickness of the colliding nuclei is related to their surface diffuseness, and can be probed by, e.g., ratios of observables in isobaric collision systems~\cite{Li:2019kkh}. Since relativistic heavy-ion collisions experience different evolution stages such as the dynamics of the QGP, the hadronization process, and the hadronic afterburner, the above observables proposed at midrapidities suffer from theoretical uncertainties of modeling these complicated dynamics. On the other hand, one may turn around to use the particle production from the spectator matter, which is not affected by the complicated dynamics in the participant region and can be measured through the ZDC, to probe the structure of colliding nuclei in relativistic heavy-ion collisions. In recent studies by some of us, we propose that the free spectator nucleons in ultracentral relativistic heavy-ion collisions can be a robust probe of the neutron-skin thickness of the colliding nuclei~\cite{Liu:2022kvz,Liu:2022xlm,Liu:2023qeq}.

Besides the deformation and the neutron skin, configurations of $\alpha$ clusters in finite nuclei have been hot topics in nuclear structures for decades~\cite{Brink:1970ufk,Bauhoff:1984zza,Freer:2017gip,Tohsaki:2017hen,BIJKER2020103735,Zhou:2019hor,Li:2020exz}. While $\alpha$ clusters may exist in the nucleus surface and influence the neutron-skin thickness~\cite{Typel:2014tqa}, the formation probability is very small in heavy nuclei, and it is expected to have minor effects on the yield of free spectator nucleons in their collisions. On the other hand, the typical $3-\alpha$ cluster and $4-\alpha$ cluster structure in $^{12}$C and $^{16}$O, respectively, have been under hot investigation for a long time~\cite{Freer:2017gip,BIJKER2020103735,Tohsaki:2017hen}, and can be related to various important questions such as the CNO cycle~\cite{Adelberger:2010qa,Wiescher:2010zz} and the nucleosynthesis of light nuclei~\cite{Coc:2011az}. $\alpha$ clusters in these light nuclei may have different configurations, e.g., linear-chain and triangle configurations in $^{12}$C, and linear-chain, tetrahedron, square, and Y-shape configurations in $^{16}$O. Various probes have been proposed to identify these configurations, e.g., the deexcitation spectra of isovector giant dipole resonances~\cite{He:2014iqa}, emissions of nucleons in photonuclear reactions~\cite{Huang:2017ysr}, direct photon production in nuclear reactions~\cite{Shi:2021far}, and anisotropic flows in relativistic heavy-ion collisions~\cite{PhysRevC.95.064904,PhysRevC.97.034912,Behera:2023nwj,PhysRevC.104.L041901}.

In the present study, we propose to probe the configuration of $\alpha$ clusters with spectator particle yields in relativistic $^{12}$C+$^{12}$C and $^{16}$O+$^{16}$O collisions. The density distributions for different $\alpha$-cluster configurations are obtained by using the microscopic cluster model with Brink wave function~\cite{Brink:1970ufk}. The participant and spectator matter are determined by the Glauber model, while a multifragmentation process is applied to the spectator matter to obtain the particle yield in the forward-backward region. While neutrons can be measured by the ZDC, charged spectator particles are deflected by the beam optics. However, these charged spectator particles are still measurable by instrumenting the forward region with dedicated detectors~\cite{Tarafdar:2014oua}. Intuitively, one may expect that more spectator $\alpha$ particles will be detected in collisions of nuclei with $\alpha$ clusters, while these $\alpha$ particles are not distinguishable by the detector from other particles with the same mass-to-charge ratio $A/Z$, such as deuterons. Considering that ZDC measures the energy deposition, i.e., counting the number of constituent nucleons, we propose that the yield ratio of free spectator neutrons to spectator particles with $A/Z=2$ scaled by their constituent nucleon numbers in ultracentral relativistic $^{12}$C+$^{12}$C and $^{16}$O+$^{16}$O collisions may serve as a useful probe of the $\alpha$-cluster configuration in these colliding nuclei.



To calculate the density distribution with $\alpha$ clusters in $^{12}$C and $^{16}$O, we adopt the following Hamiltonian
\begin{equation}
  \hat{H} =
  \sum^A_{i=1} \ E_i^{\textup{cm}}
  + \sum_{i<j} \ V^{NN}(\bm{r}_{ij})
  + \sum_{i<j} \ V^{Cou}(\bm{r}_{ij}).
  \label{eq:Hamiltonian}
\end{equation}
The summation in the above Hamiltonian is over the total nucleon number $A$. The first term represents the kinetic energy in the center-of-mass (c.m.) frame, the second term is Volkov No.2 force~\cite{Volkov_1965_NP} representing the effective nucleon-nucleon interaction, and the third term is the Coulomb interaction, with $\bm{r}_{ij}$ being the relative coordinates between nucleon $i$ and nucleon $j$. The form of the nucleon-nucleon interaction can be expressed as
\begin{equation}
  V^{NN}(\bm{r}_{ij})
  = ( V_1 e^{-\alpha_1 \bm{r}^2_{ij}} - V_2 e^{-\alpha_2 \bm{r}^2_{ij}} )
  ( W - M \hat{P}_{\sigma} \hat{P}_{\tau} + B \hat{P}_{\sigma} - H \hat{P}_{\tau} ),
  \label{eq:Volkov}
\end{equation}
where $\hat{P}_{\sigma}$ and $\hat{P}_{\tau}$ are the spin and isospin exchange operator, respectively, and $V_1 = -60.650 \ \textup{MeV}$, $V_2 = 61.140 \ \textup{MeV}$, $\alpha_1 = 0.980 \ \textup{fm}^{-2}$,  $\alpha_2 = 0.309 \ \textup{fm}^{-2}$, $W = 0.4$, $M = 0.6$, and $B = H = 0.125$ are determined from the phase shift data of $\alpha$-nucleon and $\alpha-\alpha$ scatterings as well as the binding energy of deuteron~\cite{Volkov_1965_NP, Itagaki_2000_PRC}.

We express the Brink wave function of the $n$-$\alpha$ system as
\begin{equation}
  \Phi^{\textup{Brink}}_{n \alpha} (\bm{R}_1, \cdots, \bm{R}_n )
  = n_0 \mathcal{A} \{ \psi_{\alpha_1}(\bm{R}_1) \cdots \psi_{\alpha_n}(\bm{R}_n) \},
\end{equation}
where $n_0=\sqrt{(4!)^n/A!}$ is the normalization constant, $\mathcal{A}$ denotes the antisymmetrization operator which exchanges nucleons belonging to different $\alpha$ clusters, and $\bm{R}_i$ represents the center of the $i$th cluster, with its wave function expressed as
\begin{equation}
  \psi_{\alpha_i}(\bm{R}_i)
  = n_1 \det \{ \phi_{1}(\bm{r}_1-\bm{R}_i) \cdots \phi_{4}(\bm{r}_4-\bm{R}_i) \}.
\end{equation}
In the above, $n_1=1/\sqrt{4!}$ is the normalization constant, $\det$ represents the Slater determinant, and $\phi(\bm{r}-\bm{R}_i)$ denotes the single-nucleon wave function
\begin{equation}
  \phi(\bm{r}-\bm{R}_i)
  = (\pi b^2)^{-3/4} \exp\left[-\frac{(\bm{r}-\bm{R}_i)^2}{2 b^2}\right] \chi(\sigma, \tau).
\end{equation}
Here, the spatial part is defined as the $1s$ harmonic-oscillator orbit centered around $\bm{R}_i$, with the size parameter $b=1.46$ fm for each nucleon~\cite{Itagaki_1995_PTP}, and $\chi(\sigma, \tau)$ denotes the spin and isospin part.

The above Brink wave function contains the intrinsic wave function $\Phi^{\textup{Int}}$ and the wave function $\Phi^{\textup{cm}}$ due to the c.m. motion, i.e.,
\begin{equation}
  \Phi^{\textup{Brink}} = \Phi^{\textup{Int}} \Phi^{\textup{cm}},
  \label{eq:wave-function-two-parts}
\end{equation}
where $\Phi^{\textup{cm}}$ can be expressed as
\begin{equation}
  \Phi^{\textup{cm}}
  = \left(\frac{A}{\pi b^2}\right)^{3/4} \textup{exp}\left( -\frac{A}{2 b^2} \bm{X}^2_{\textup{cm}} \right),
\end{equation}
with $\bm{X}_{\textup{cm}} = \frac{1}{A} \sum^A_{i=1} \bm{r}_i$ being the c.m. coordinate.
To restore the rotational symmetry, we project the Brink wave function through
\begin{equation}
  \Psi^{J}_{M}
  = \hat{\mathcal{P}}^{J}_{M,K} \Phi^{\textup{Brink}},
  \label{eq:total_wave_function}
\end{equation}
where the corresponding angular momentum projection operator is expressed as~\cite{Ring_2004_Book}
\begin{eqnarray}
  \hat{\mathcal{P}}^{J}_{M,K} =
  \frac{2J+1}{16\pi^2} \ \int^{2\pi}_0 d\phi \
  \int^{\pi}_0 d \theta\ \mathrm{sin}(\theta) \notag \\
 \times \int^{4\pi}_0 d\gamma \
  D^{J*}_{M,K}(\phi, \theta, \gamma)
  \hat{R}(\phi, \theta, \gamma),
  \label{eq:angular_momentum_projection}
\end{eqnarray}
with $D^{J}_{M,K}(\phi, \theta, \gamma)$ being the Wigner rotation matrix, $\hat{R}$ being the rotation operator, and $\phi$, $\theta$, and $\gamma$ being the three Euler angles. $J$ and $M$ are respectively the quantum number of the total angular momentum and that in the third direction after projection, $K$ is the one before projection, and they are all zero for $^{12}$C and $^{16}$O in the present study. In order to get the wave function and energy of the ground state, the variational principle is used, and the distance between nucleons for a specific cluster configuration is determined by finding the minimum energy based on $\Psi^{J}_{M}$. With the structure parameters determined, the density at position $\bm{a}$ is then calculated based on the intrinsic wave function $\Phi^{\textup{Int}}$ without the angular momentum projection through
\begin{eqnarray}
  \rho(\bm{a})
  &=& \sum_{i=1}^A\frac{\langle \Phi^{\text{Int}} | \delta(\bm{r}_i - \bm{a}) | \Phi^{\text{Int}} \rangle}{\langle \Phi^{\text{Int}} | \Phi^{\text{Int}} \rangle} \notag\\
  &=& \sum_{i=1}^A \frac{1}{(2 \pi)^3} \int d^3 \bm{k} \ e^{-i \bm{k} \cdot \bm{a}}
  \frac{\langle \Phi^{\text{Int}} | e^{i \bm{k} \cdot \bm{r}_i} | \Phi^{\text{Int}} \rangle}
  {\langle \Phi^{\text{Int}} | \Phi^{\text{Int}} \rangle}.
\end{eqnarray}

\begin{figure}[ht]
  \centering
  \includegraphics[scale=0.4]{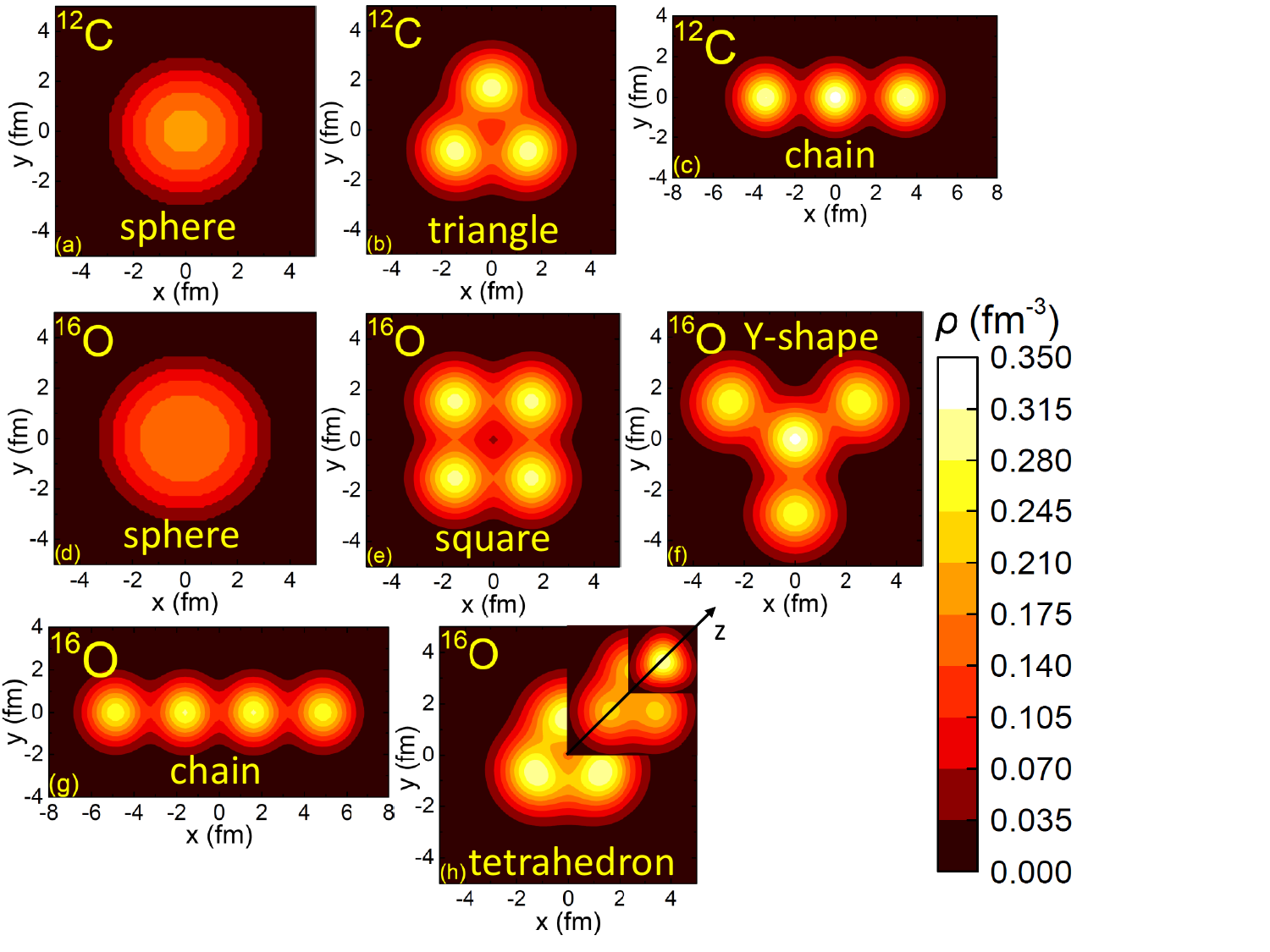}
  \caption{Density distributions for different configurations of $\alpha$ clusters in the x-o-y plane, i.e., sphere (no $\alpha$ clusters) (a), triangle (b), and chain (c) configurations for $^{12}$C, and sphere (no $\alpha$ clusters) (d), square (e), Y-shape (f), chain (g), and tetrahedron (h) configurations for $^{16}$O. }
  \label{fig:density_dis}
\end{figure}

Density distributions for different configurations of $^{12}$C and $^{16}$O are displayed in Fig.~\ref{fig:density_dis}. We have considered configurations with $\alpha$ clusters of triangle and chain structures in the 3-$\alpha$ $^{12}$C system, and the square, Y-shape, chain, and tetrahedron structures in the 4-$\alpha$ $^{16}$O system. We have also considered the sphere case without $\alpha$ clusters, and the corresponding density distribution is obtained from the Skyrme-Hartree-Fock (SHF) calculation by using the MSL0 force~\cite{Chen:2010qx}. As shown in Table \ref{tab:RMS}, the compact $\alpha$-cluster configurations reproduce the experimental data of charge radii $R_{ch}=2.47$ fm for $^{12}$C and $R_{ch}=2.70$ fm for $^{16}$O~\cite{Angeli:2004kvy} reasonably well.

\begin{table}[h!]
\centering
\caption{Root-mean-square (RMS) radii in fm of $^{12}$C and $^{16}$O for different configurations.}
\label{tab:RMS}
\renewcommand\arraystretch{1.5}
\setlength{\tabcolsep}{1.0mm}
\begin{tabular}{cccccccc}
\hline\hline
       &  sphere   & triangle & square & chain & Y-shape & tetrahedron  \\
\hline
 $^{12}$C & 2.49& 2.45  & -  & 3.29 & -  & -  \\
\hline
 $^{16}$O & 2.69& - &  2.77  & 4.06 & 3.09  & 2.45  \\
\hline\hline
\end{tabular}
\end{table}

\begin{figure}[ht]
\includegraphics[width=1.\linewidth]{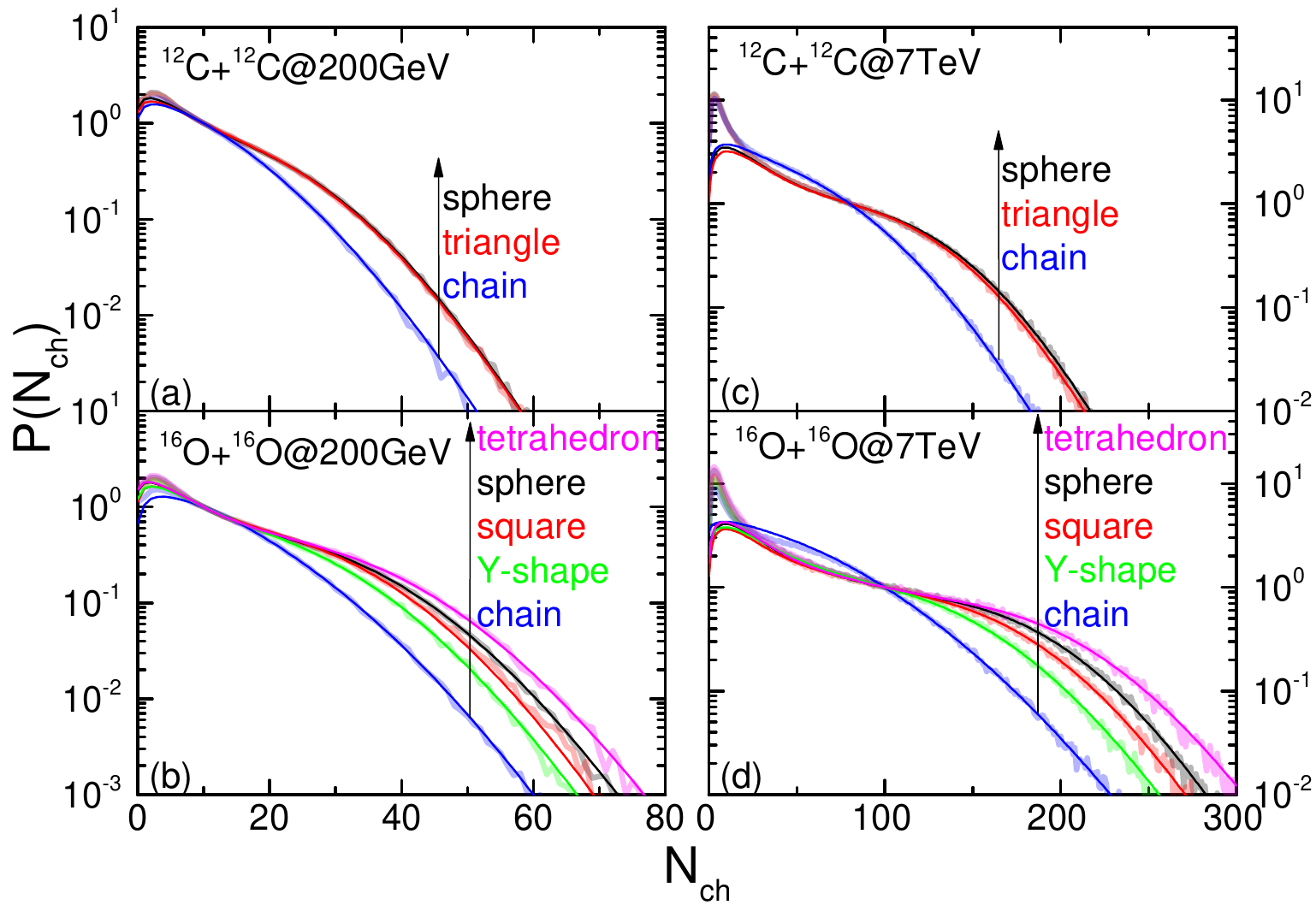}
\vspace*{-.5cm}
\caption{\label{fig:Nchdis} (Color online) Rescaled distributions of mid-rapidity charged-particle multiplicity $\nch$ in $^{12}$C+$^{12}$C (upper) and $^{16}$O+$^{16}$O (lower) collisions at \sqrtsnn = 200 GeV (left) and 7 TeV (right) with different density distributions shown in Fig.~\ref{fig:density_dis}. The bands represent results from AMPT calculations, and they can be fitted by the solid lines based on the Glauber model.}
\end{figure}

We try to distinguish different configurations shown above by using spectator particle yields in relativistic heavy-ion collisions. To be comparable to future experimental data, the results for different configurations will be compared at the same centrality range. Due to the lacking of the experimental data so far, we generate the distribution of charged-particle multiplicity ($\nch$) in minibias collisions from a multiphase transport model (AMPT)~\cite{Lin:2004en}, to determine the relation between the impact parameter and the $\nch$ or the centrality. In different collision systems and taking different density distributions from Fig.~\ref{fig:density_dis} with randomized orientations as input, the obtained $\nch$ distributions within the pseudorapidity acceptance $|\eta|<0.5$ are shown in Fig.~\ref{fig:Nchdis} with bands. It is seen that a more compact density distribution (or a smaller RMS radius as seen in Table~\ref{tab:RMS}), such as sphere and triangle for $^{12}$C as well as sphere and tetrahedron for $^{16}$O, is likely to generate events with large $\nch$, while it is difficult for the chain configurations in both nuclei to generate events with large $\nch$. This is intuitively understandable, since the number of participant nucleons is generally larger (smaller) for a more compact (expanded) configuration. The obtained $\nch$ distributions can be fitted based on the two-component Glauber model~\cite{Miller:2007ri,Kharzeev:2000ph}. In this way, the results can be presented as a function of $\nch$, or equivalently the centrality, as in the experimental analysis.


While the information of the spectator matter can in principle also be obtained from the AMPT model, a separate framework by neglecting the participant dynamics is performed in order to speed up the calculation. In this framework, we first sample the coordinates of initial neutrons and protons in colliding nuclei according to the density distributions in Fig.~\ref{fig:density_dis}, while nucleon momenta are sampled isotropically within the Fermi sphere, with the Fermi momentum calculated according to the local nucleon density. By using the nucleon-nucleon inelastic cross section of 42 mb at \sqrtsnn = 200 GeV and 71 mb at \sqrtsnn = 7 TeV~\cite{Loizides:2017ack}, a Monte-Carlo Glauber model~\cite{Miller:2007ri} is then used to determine spectator nucleons which do not experience any nucleon-nucleon collisions. The spectator nucleons are further grouped into heavy clusters ($A \geq 4$) and free nucleons based on a minimum spanning tree algorithm, i.e., nucleons with their distance $\Delta r<\Delta r_{\mathrm{max}}$ and relative momentum $\Delta p<p_{\mathrm{max}}$ may form heavy clusters. The coalescence parameters $\Delta r_{\mathrm{max}}=3$~fm and $\Delta p_{\mathrm{max}}=300$ MeV/$c$ taken from Ref.~\cite{Li:1997rc} have been shown to give the best description of the experimental data of free spectator neutrons in ultracentral \auau\ collisions at $\sqrtsnn=130$ GeV~\cite{Liu:2022kvz}. For spectator nucleons that do not form heavy clusters ($A \geq 4$), they may still have chance to coalesce into light clusters with $A \leq 3$, i.e., deuterons, tritons, and $^3$He, and this process is implemented based on a Wigner function approach~\cite{Chen:2003ava,Sun:2017ooe}. The deexcitation of heavy clusters with $A \geq 4$ are handled by the GEMINI model~\cite{Charity:1988zz,Charity:2010wk}, which requires as inputs the angular momentum and the excitation energy of the cluster. The angular momentum of the cluster is calculated by summing those from all nucleons with respective to their center of mass. The energy of the cluster is calculated from a simplified SHF energy-density functional~\cite{Chen:2010qx} based on the neutron and proton phase-space information obtained from the test-particle method~\cite{Wong:1982zzb,Bertsch:1988ik}, and its excitation energy is then calculated by subtracting from the calculated cluster energy the ground-state energy taken from the mass table~\cite{Wang:2021xhn} or an improved liquid-drop model~\cite{Wang:2014qqa}. For more details of the above multifragmentation framework for the spectator matter, we refer the reader to Refs.~\cite{Liu:2022kvz,Liu:2022xlm}.

\begin{figure}[ht]
\includegraphics[width=1.\linewidth]{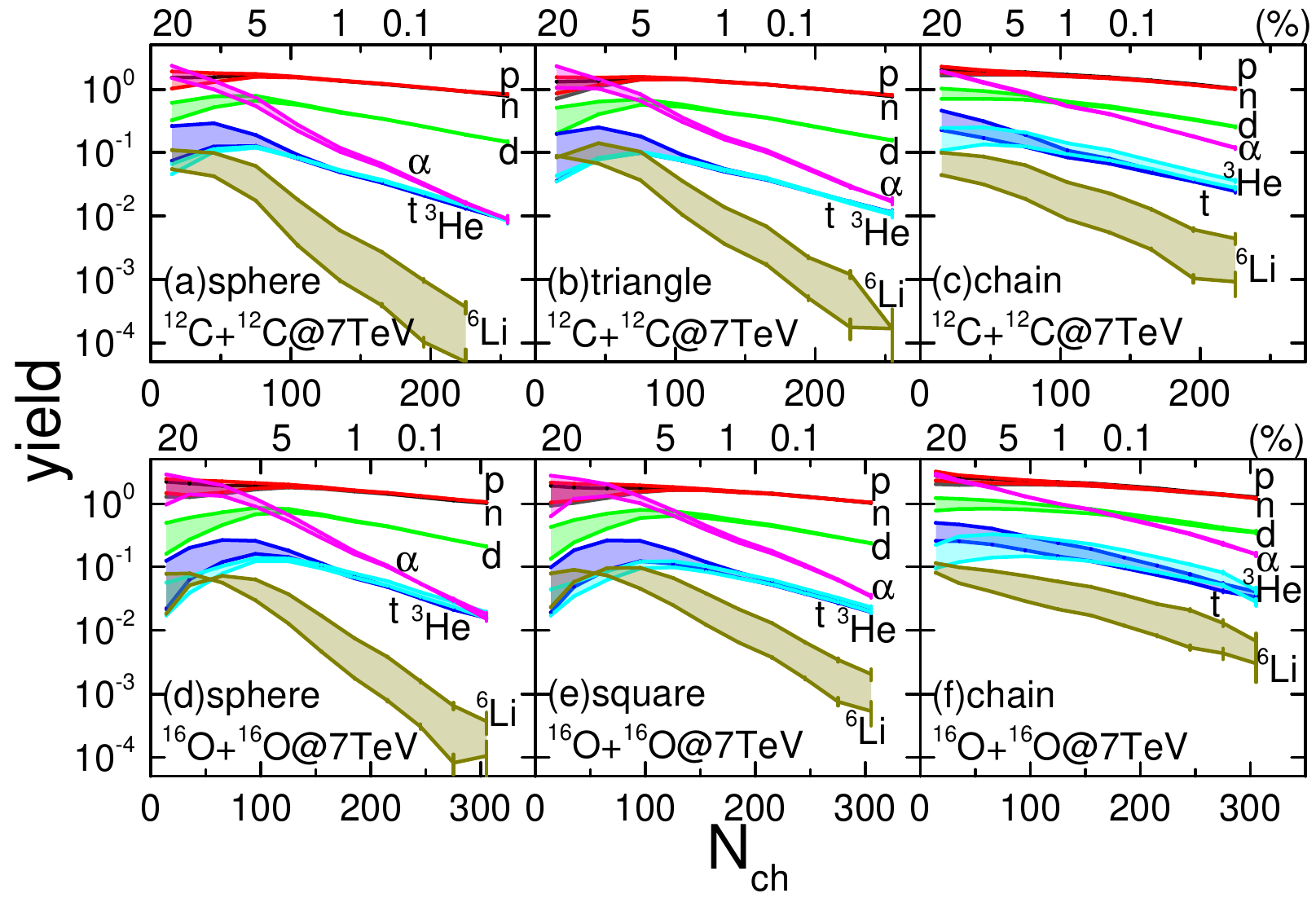}
\vspace*{-.5cm}
\caption{\label{fig:yield} (Color online) Charged-particle multiplicity $N_{ch}$ or centrality dependence of spectator particle yields in $^{12}$C+$^{12}$C collisions with sphere, triangle, and chain configurations, and $^{16}$O+$^{16}$O collisions with sphere, square, and chain configurations, at $\sqrtsnn$ = 7 TeV. }
\end{figure}

With the theoretical framework described above, we can get the yields of spectator particles such as neutrons, protons, deuterons, tritons, $^3$He, $\alpha$ particles, etc., and typical results from sphere, triangle, chain configurations for $^{12}$C+$^{12}$C collisions and those from sphere, square, and chain configurations for $^{16}$O+$^{16}$O collisions at \sqrtsnn = 7 TeV are shown in Fig.~\ref{fig:yield}. For free spectator neutrons and protons, they are composed of the residue ones from direct production that have not coalesced into light clusters and those from the deexcitation of heavy clusters by GEMINI. For spectator deuterons, tritons, and $^3$He, they are generated from the coalescence of directly produced neutrons and protons as well as from the deexcitation of heavy clusters by GEMINI. For $\alpha$ particles or even heavier clusters, they are mostly produced from the deexcitation of heavy clusters by GEMINI as well as the stable ones from the clusterization algorithm. The bands in Fig.~\ref{fig:yield} represent the uncertainty of $\pm1$ MeV per nucleon in calculating the excitation energy for the deexcitation of heavy fragments by GEMINI, estimated from the same degree of accuracy for calculating the ground-state energy of the relevant nucleus from the nucleon phase-space information compared to the microscopic model. The uncertainties in the deexcitation process are seen to be largely reduced in ultracentral collisions.

\textbf{}

The sphere and triangle configurations for $^{12}$C as well as the sphere and square configurations for $^{16}$O have the similar RMS radii, as can be seen from Table~\ref{tab:RMS}, so the average spectator nucleon numbers are expected to be similar. Comparing Fig.~\ref{fig:yield}(a) with Fig.~\ref{fig:yield}(b) and Fig.~\ref{fig:yield}(d) with Fig.~\ref{fig:yield}(e), we can see that the triangle configuration for $^{12}$C and the square configuration for $^{16}$O lead to a larger $\alpha$ particle yield and also slightly larger yields for deuterons, tritons, and $^3$He, compared to the corresponding sphere configuration, since the triangle or square configuration has initial $\alpha$ clusters, so neutrons and protons in spectator matter are close in phase space and are more likely to form $\alpha$ clusters or other light nuclei. Comparing Fig.~\ref{fig:yield}(b) with Fig.~\ref{fig:yield}(c) and Fig.~\ref{fig:yield}(e) with Fig.~\ref{fig:yield}(f), we can see that the chain configuration, which has a larger RMS radius and deformation, leads to a larger spectator nucleon number and thus larger yields for all particle species at the same centrality, especially for $\alpha$ particles, compared to the triangle or square configuration.

\begin{figure}[htb]
\includegraphics[width=1.\linewidth]{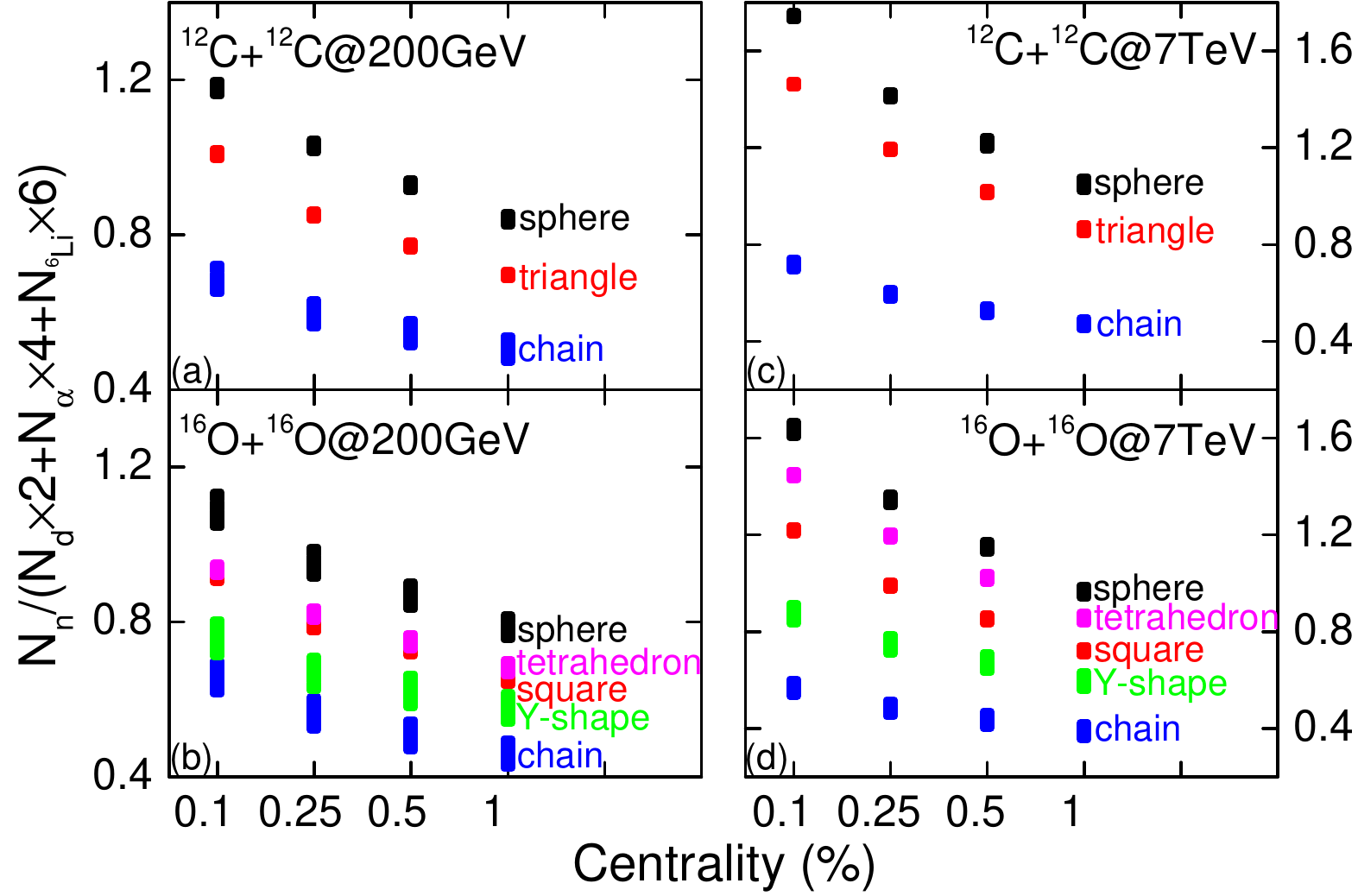}
\vspace*{-.5cm}
\caption{\label{fig:yield_ratio} (Color online) Yield ratio of free spectator neutrons to charged particles with $A/Z=2$ scaled by their constituent nucleon numbers in ultracentral $^{12}$C+$^{12}$C (upper) and $^{16}$O+$^{16}$O (lower) collisions at $\sqrtsnn$ = 200 GeV (left) and 7 TeV (right).}
\end{figure}

To reduce the uncertainty of absolute particle yields from both theoretical and experimental sides, we propose to take ratios of spectator particle yields to probe different $\alpha$-cluster configurations in colliding nuclei. Since particles with the same $A/Z$ are affected by the beam optics in the same way, we propose to use the yield ratio $N_n/N_{A/Z=2}$ of free spectator neutrons to spectator particles with $A/Z=2$ scaled by their constituent nucleon numbers, and the results for ultracentral $^{12}$C+$^{12}$C and $^{16}$O+$^{16}$O collisions at RHIC and LHC energies are shown in Fig.~\ref{fig:yield_ratio}, where the height of bars represents again the uncertainty due to the deexcitation process (estimated from $\pm1$ MeV per nucleon in the excitation energy). As mentioned above, collisions of nuclei without initial $\alpha$ clusters  generally generate less spectator light nuclei than collisions of nuclei with initial $\alpha$ clusters at the same centrality, and therefore leads to the largest $N_n/N_{A/Z=2}$ ratio. For collisions of nuclei with initial chain structure of $\alpha$ clusters, they lead to a more massive spectator matter, which generates more spectator free neutrons but even more light nuclei, as shown in Fig.~\ref{fig:yield}, and thus the smallest $N_n/N_{A/Z=2}$ ratio. For $N_n/N_{A/Z=2}$ ratios from other configurations, they lie in between the above two extreme cases, and the values depend on their RMS radii shown in Table~\ref{tab:RMS}. At LHC energy $\sqrtsnn=7$ TeV shown in the right panels of Fig.~\ref{fig:yield_ratio}, the larger nucleon-nucleon collision cross section leads to a smaller spectator nucleon number at the same centrality, and this generally results in a larger $N_n/N_{A/Z=2}$ ratio, compared to that at RHIC energy $\sqrtsnn=200$ GeV. The larger cross section at LHC energy also reduces the uncertainty of the deexcitation process, thus helps to probe different $\alpha$-cluster configurations more clearly.

To summarize, we propose to use the yield ratio $N_n/N_{A/Z=2}$ of free spectator neutrons to spectator particles with mass-to-charge ratio $A/Z=2$ scaled by their masses in ultracentral relativistic $^{12}$C+$^{12}$C and $^{16}$O+$^{16}$O collisions as a sensitive probe of the $\alpha$-cluster structure in $^{12}$C and $^{16}$O. The idea is illustrated based on initial density distributions with different $\alpha$-cluster structures generated by a microscopic cluster model with Brink wave function and without $\alpha$ clusters generated by the SHF calculation. The multifragmentation of the spectator matter generates more light nuclei for collisions of nuclei with initial $\alpha$ clusters compared to those of nuclei without initial $\alpha$ clusters, and collisions of nuclei with a chain structure of $\alpha$ clusters lead to the largest yield of spectator light nuclei. Therefore, the $N_n/N_{A/Z=2}$ ratio is smallest for the chain structure but larger for more compact configurations. Collisions at higher energies, such as those at LHC, lead to a larger nucleon-nucleon cross section, and may help to probe different $\alpha$-cluster configurations more clearly.


This work is supported by the Strategic Priority Research Program of the Chinese Academy of Sciences under Grant No. XDB34030000, the National Natural Science Foundation of China under Grant Nos. 12375125, 12035011, 11975167, 12147101, 12225502, and 12075061, the National Key Research and Development Program of China under Grant No. 2022YFA1604900, and the Fundamental Research Funds for the Central Universities.

\bibliography{alphacluster}

\begin{thebibliography}{46}%
\makeatletter
\providecommand \@ifxundefined [1]{%
 \@ifx{#1\undefined}
}%
\providecommand \@ifnum [1]{%
 \ifnum #1\expandafter \@firstoftwo
 \else \expandafter \@secondoftwo
 \fi
}%
\providecommand \@ifx [1]{%
 \ifx #1\expandafter \@firstoftwo
 \else \expandafter \@secondoftwo
 \fi
}%
\providecommand \natexlab [1]{#1}%
\providecommand \enquote  [1]{``#1''}%
\providecommand \bibnamefont  [1]{#1}%
\providecommand \bibfnamefont [1]{#1}%
\providecommand \citenamefont [1]{#1}%
\providecommand \href@noop [0]{\@secondoftwo}%
\providecommand \href [0]{\begingroup \@sanitize@url \@href}%
\providecommand \@href[1]{\@@startlink{#1}\@@href}%
\providecommand \@@href[1]{\endgroup#1\@@endlink}%
\providecommand \@sanitize@url [0]{\catcode `\\12\catcode `\$12\catcode
  `\&12\catcode `\#12\catcode `\^12\catcode `\_12\catcode `\%12\relax}%
\providecommand \@@startlink[1]{}%
\providecommand \@@endlink[0]{}%
\providecommand \url  [0]{\begingroup\@sanitize@url \@url }%
\providecommand \@url [1]{\endgroup\@href {#1}{\urlprefix }}%
\providecommand \urlprefix  [0]{URL }%
\providecommand \Eprint [0]{\href }%
\providecommand \doibase [0]{http://dx.doi.org/}%
\providecommand \selectlanguage [0]{\@gobble}%
\providecommand \bibinfo  [0]{\@secondoftwo}%
\providecommand \bibfield  [0]{\@secondoftwo}%
\providecommand \translation [1]{[#1]}%
\providecommand \BibitemOpen [0]{}%
\providecommand \bibitemStop [0]{}%
\providecommand \bibitemNoStop [0]{.\EOS\space}%
\providecommand \EOS [0]{\spacefactor3000\relax}%
\providecommand \BibitemShut  [1]{\csname bibitem#1\endcsname}%
\let\auto@bib@innerbib\@empty
\bibitem [{\citenamefont {Filip}\ \emph {et~al.}(2009)\citenamefont {Filip},
  \citenamefont {Lednicky}, \citenamefont {Masui},\ and\ \citenamefont
  {Xu}}]{Filip:2009zz}%
  \BibitemOpen
  \bibfield  {author} {\bibinfo {author} {\bibfnamefont {P.}~\bibnamefont
  {Filip}}, \bibinfo {author} {\bibfnamefont {R.}~\bibnamefont {Lednicky}},
  \bibinfo {author} {\bibfnamefont {H.}~\bibnamefont {Masui}}, \ and\ \bibinfo
  {author} {\bibfnamefont {N.}~\bibnamefont {Xu}},\ }\href {\doibase
  10.1103/PhysRevC.80.054903} {\bibfield  {journal} {\bibinfo  {journal} {Phys.
  Rev. C}\ }\textbf {\bibinfo {volume} {80}},\ \bibinfo {pages} {054903}
  (\bibinfo {year} {2009})}\BibitemShut {NoStop}%
\bibitem [{\citenamefont {Shou}\ \emph {et~al.}(2015)\citenamefont {Shou},
  \citenamefont {Ma}, \citenamefont {Sorensen}, \citenamefont {Tang},
  \citenamefont {Videb\ae{}k},\ and\ \citenamefont {Wang}}]{Shou:2014eya}%
  \BibitemOpen
  \bibfield  {author} {\bibinfo {author} {\bibfnamefont {Q.~Y.}\ \bibnamefont
  {Shou}}, \bibinfo {author} {\bibfnamefont {Y.~G.}\ \bibnamefont {Ma}},
  \bibinfo {author} {\bibfnamefont {P.}~\bibnamefont {Sorensen}}, \bibinfo
  {author} {\bibfnamefont {A.~H.}\ \bibnamefont {Tang}}, \bibinfo {author}
  {\bibfnamefont {F.}~\bibnamefont {Videb\ae{}k}}, \ and\ \bibinfo {author}
  {\bibfnamefont {H.}~\bibnamefont {Wang}},\ }\href {\doibase
  10.1016/j.physletb.2015.07.078} {\bibfield  {journal} {\bibinfo  {journal}
  {Phys. Lett. B}\ }\textbf {\bibinfo {volume} {749}},\ \bibinfo {pages} {215}
  (\bibinfo {year} {2015})},\ \Eprint {http://arxiv.org/abs/1409.8375}
  {arXiv:1409.8375 [nucl-th]} \BibitemShut {NoStop}%
\bibitem [{\citenamefont {Giacalone}(2020)}]{Giacalone:2019pca}%
  \BibitemOpen
  \bibfield  {author} {\bibinfo {author} {\bibfnamefont {G.}~\bibnamefont
  {Giacalone}},\ }\href {\doibase 10.1103/PhysRevLett.124.202301} {\bibfield
  {journal} {\bibinfo  {journal} {Phys. Rev. Lett.}\ }\textbf {\bibinfo
  {volume} {124}},\ \bibinfo {pages} {202301} (\bibinfo {year} {2020})},\
  \Eprint {http://arxiv.org/abs/1910.04673} {arXiv:1910.04673 [nucl-th]}
  \BibitemShut {NoStop}%
\bibitem [{\citenamefont {Zhang}\ and\ \citenamefont
  {Jia}(2022)}]{Zhang:2021kxj}%
  \BibitemOpen
  \bibfield  {author} {\bibinfo {author} {\bibfnamefont {C.}~\bibnamefont
  {Zhang}}\ and\ \bibinfo {author} {\bibfnamefont {J.}~\bibnamefont {Jia}},\
  }\href {\doibase 10.1103/PhysRevLett.128.022301} {\bibfield  {journal}
  {\bibinfo  {journal} {Phys. Rev. Lett.}\ }\textbf {\bibinfo {volume} {128}},\
  \bibinfo {pages} {022301} (\bibinfo {year} {2022})},\ \Eprint
  {http://arxiv.org/abs/2109.01631} {arXiv:2109.01631 [nucl-th]} \BibitemShut
  {NoStop}%
\bibitem [{\citenamefont {Li}\ \emph {et~al.}(2020{\natexlab{a}})\citenamefont
  {Li}, \citenamefont {Xu}, \citenamefont {Zhou}, \citenamefont {Wang},
  \citenamefont {Zhao}, \citenamefont {Chen},\ and\ \citenamefont
  {Wang}}]{Li:2019kkh}%
  \BibitemOpen
  \bibfield  {author} {\bibinfo {author} {\bibfnamefont {H.}~\bibnamefont
  {Li}}, \bibinfo {author} {\bibfnamefont {H.-j.}\ \bibnamefont {Xu}}, \bibinfo
  {author} {\bibfnamefont {Y.}~\bibnamefont {Zhou}}, \bibinfo {author}
  {\bibfnamefont {X.}~\bibnamefont {Wang}}, \bibinfo {author} {\bibfnamefont
  {J.}~\bibnamefont {Zhao}}, \bibinfo {author} {\bibfnamefont {L.-W.}\
  \bibnamefont {Chen}}, \ and\ \bibinfo {author} {\bibfnamefont
  {F.}~\bibnamefont {Wang}},\ }\href {\doibase 10.1103/PhysRevLett.125.222301}
  {\bibfield  {journal} {\bibinfo  {journal} {Phys. Rev. Lett.}\ }\textbf
  {\bibinfo {volume} {125}},\ \bibinfo {pages} {222301} (\bibinfo {year}
  {2020}{\natexlab{a}})},\ \Eprint {http://arxiv.org/abs/1910.06170}
  {arXiv:1910.06170 [nucl-th]} \BibitemShut {NoStop}%
\bibitem [{\citenamefont {Liu}\ \emph {et~al.}(2022{\natexlab{a}})\citenamefont
  {Liu}, \citenamefont {Zhang}, \citenamefont {Zhou}, \citenamefont {Xu},
  \citenamefont {Jia},\ and\ \citenamefont {Peng}}]{Liu:2022kvz}%
  \BibitemOpen
  \bibfield  {author} {\bibinfo {author} {\bibfnamefont {L.-M.}\ \bibnamefont
  {Liu}}, \bibinfo {author} {\bibfnamefont {C.-J.}\ \bibnamefont {Zhang}},
  \bibinfo {author} {\bibfnamefont {J.}~\bibnamefont {Zhou}}, \bibinfo {author}
  {\bibfnamefont {J.}~\bibnamefont {Xu}}, \bibinfo {author} {\bibfnamefont
  {J.}~\bibnamefont {Jia}}, \ and\ \bibinfo {author} {\bibfnamefont {G.-X.}\
  \bibnamefont {Peng}},\ }\href {\doibase 10.1016/j.physletb.2022.137441}
  {\bibfield  {journal} {\bibinfo  {journal} {Phys. Lett. B}\ }\textbf
  {\bibinfo {volume} {834}},\ \bibinfo {pages} {137441} (\bibinfo {year}
  {2022}{\natexlab{a}})},\ \Eprint {http://arxiv.org/abs/2203.09924}
  {arXiv:2203.09924 [nucl-th]} \BibitemShut {NoStop}%
\bibitem [{\citenamefont {Liu}\ \emph {et~al.}(2022{\natexlab{b}})\citenamefont
  {Liu}, \citenamefont {Zhang}, \citenamefont {Xu}, \citenamefont {Jia},\ and\
  \citenamefont {Peng}}]{Liu:2022xlm}%
  \BibitemOpen
  \bibfield  {author} {\bibinfo {author} {\bibfnamefont {L.-M.}\ \bibnamefont
  {Liu}}, \bibinfo {author} {\bibfnamefont {C.-J.}\ \bibnamefont {Zhang}},
  \bibinfo {author} {\bibfnamefont {J.}~\bibnamefont {Xu}}, \bibinfo {author}
  {\bibfnamefont {J.}~\bibnamefont {Jia}}, \ and\ \bibinfo {author}
  {\bibfnamefont {G.-X.}\ \bibnamefont {Peng}},\ }\href {\doibase
  10.1103/PhysRevC.106.034913} {\bibfield  {journal} {\bibinfo  {journal}
  {Phys. Rev. C}\ }\textbf {\bibinfo {volume} {106}},\ \bibinfo {pages}
  {034913} (\bibinfo {year} {2022}{\natexlab{b}})},\ \Eprint
  {http://arxiv.org/abs/2209.03106} {arXiv:2209.03106 [nucl-th]} \BibitemShut
  {NoStop}%
\bibitem [{\citenamefont {Liu}\ \emph {et~al.}(2023)\citenamefont {Liu},
  \citenamefont {Xu},\ and\ \citenamefont {Peng}}]{Liu:2023qeq}%
  \BibitemOpen
  \bibfield  {author} {\bibinfo {author} {\bibfnamefont {L.-M.}\ \bibnamefont
  {Liu}}, \bibinfo {author} {\bibfnamefont {J.}~\bibnamefont {Xu}}, \ and\
  \bibinfo {author} {\bibfnamefont {G.-X.}\ \bibnamefont {Peng}},\ }\href
  {\doibase 10.1016/j.physletb.2023.137701} {\bibfield  {journal} {\bibinfo
  {journal} {Phys. Lett. B}\ }\textbf {\bibinfo {volume} {838}},\ \bibinfo
  {pages} {137701} (\bibinfo {year} {2023})},\ \Eprint
  {http://arxiv.org/abs/2301.07893} {arXiv:2301.07893 [nucl-th]} \BibitemShut
  {NoStop}%
\bibitem [{\citenamefont {Brink}\ \emph {et~al.}(1970)\citenamefont {Brink},
  \citenamefont {Friedrich}, \citenamefont {Weiguny},\ and\ \citenamefont
  {Wong}}]{Brink:1970ufk}%
  \BibitemOpen
  \bibfield  {author} {\bibinfo {author} {\bibfnamefont {D.~M.}\ \bibnamefont
  {Brink}}, \bibinfo {author} {\bibfnamefont {H.}~\bibnamefont {Friedrich}},
  \bibinfo {author} {\bibfnamefont {A.}~\bibnamefont {Weiguny}}, \ and\
  \bibinfo {author} {\bibfnamefont {C.~W.}\ \bibnamefont {Wong}},\ }\href
  {\doibase 10.1016/0370-2693(70)90284-4} {\bibfield  {journal} {\bibinfo
  {journal} {Phys. Lett. B}\ }\textbf {\bibinfo {volume} {33}},\ \bibinfo
  {pages} {143} (\bibinfo {year} {1970})}\BibitemShut {NoStop}%
\bibitem [{\citenamefont {Bauhoff}\ \emph {et~al.}(1984)\citenamefont
  {Bauhoff}, \citenamefont {Schultheis},\ and\ \citenamefont
  {Schultheis}}]{Bauhoff:1984zza}%
  \BibitemOpen
  \bibfield  {author} {\bibinfo {author} {\bibfnamefont {W.}~\bibnamefont
  {Bauhoff}}, \bibinfo {author} {\bibfnamefont {H.}~\bibnamefont {Schultheis}},
  \ and\ \bibinfo {author} {\bibfnamefont {R.}~\bibnamefont {Schultheis}},\
  }\href {\doibase 10.1103/PhysRevC.29.1046} {\bibfield  {journal} {\bibinfo
  {journal} {Phys. Rev. C}\ }\textbf {\bibinfo {volume} {29}},\ \bibinfo
  {pages} {1046} (\bibinfo {year} {1984})}\BibitemShut {NoStop}%
\bibitem [{\citenamefont {Freer}\ \emph {et~al.}(2018)\citenamefont {Freer},
  \citenamefont {Horiuchi}, \citenamefont {Kanada-En'yo}, \citenamefont {Lee},\
  and\ \citenamefont {Mei\ss{}ner}}]{Freer:2017gip}%
  \BibitemOpen
  \bibfield  {author} {\bibinfo {author} {\bibfnamefont {M.}~\bibnamefont
  {Freer}}, \bibinfo {author} {\bibfnamefont {H.}~\bibnamefont {Horiuchi}},
  \bibinfo {author} {\bibfnamefont {Y.}~\bibnamefont {Kanada-En'yo}}, \bibinfo
  {author} {\bibfnamefont {D.}~\bibnamefont {Lee}}, \ and\ \bibinfo {author}
  {\bibfnamefont {U.-G.}\ \bibnamefont {Mei\ss{}ner}},\ }\href {\doibase
  10.1103/RevModPhys.90.035004} {\bibfield  {journal} {\bibinfo  {journal}
  {Rev. Mod. Phys.}\ }\textbf {\bibinfo {volume} {90}},\ \bibinfo {pages}
  {035004} (\bibinfo {year} {2018})},\ \Eprint
  {http://arxiv.org/abs/1705.06192} {arXiv:1705.06192 [nucl-th]} \BibitemShut
  {NoStop}%
\bibitem [{\citenamefont {Tohsaki}\ \emph {et~al.}(2017)\citenamefont
  {Tohsaki}, \citenamefont {Horiuchi}, \citenamefont {Schuck},\ and\
  \citenamefont {Roepke}}]{Tohsaki:2017hen}%
  \BibitemOpen
  \bibfield  {author} {\bibinfo {author} {\bibfnamefont {A.}~\bibnamefont
  {Tohsaki}}, \bibinfo {author} {\bibfnamefont {H.}~\bibnamefont {Horiuchi}},
  \bibinfo {author} {\bibfnamefont {P.}~\bibnamefont {Schuck}}, \ and\ \bibinfo
  {author} {\bibfnamefont {G.}~\bibnamefont {Roepke}},\ }\href {\doibase
  10.1103/RevModPhys.89.011002} {\bibfield  {journal} {\bibinfo  {journal}
  {Rev. Mod. Phys.}\ }\textbf {\bibinfo {volume} {89}},\ \bibinfo {pages}
  {011002} (\bibinfo {year} {2017})},\ \Eprint
  {http://arxiv.org/abs/1702.04591} {arXiv:1702.04591 [nucl-th]} \BibitemShut
  {NoStop}%
\bibitem [{\citenamefont {Bijker}\ and\ \citenamefont
  {Iachello}(2020)}]{BIJKER2020103735}%
  \BibitemOpen
  \bibfield  {author} {\bibinfo {author} {\bibfnamefont {R.}~\bibnamefont
  {Bijker}}\ and\ \bibinfo {author} {\bibfnamefont {F.}~\bibnamefont
  {Iachello}},\ }\href {\doibase https://doi.org/10.1016/j.ppnp.2019.103735}
  {\bibfield  {journal} {\bibinfo  {journal} {Progress in Particle and Nuclear
  Physics}\ }\textbf {\bibinfo {volume} {110}},\ \bibinfo {pages} {103735}
  (\bibinfo {year} {2020})}\BibitemShut {NoStop}%
\bibitem [{\citenamefont {Zhou}\ \emph {et~al.}(2019)\citenamefont {Zhou},
  \citenamefont {Funaki}, \citenamefont {Horiuchi}, \citenamefont {Kimura},
  \citenamefont {Ren}, \citenamefont {R\"opke}, \citenamefont {Schuck},
  \citenamefont {Tohsaki}, \citenamefont {Xu},\ and\ \citenamefont
  {Yamada}}]{Zhou:2019hor}%
  \BibitemOpen
  \bibfield  {author} {\bibinfo {author} {\bibfnamefont {B.}~\bibnamefont
  {Zhou}}, \bibinfo {author} {\bibfnamefont {Y.}~\bibnamefont {Funaki}},
  \bibinfo {author} {\bibfnamefont {H.}~\bibnamefont {Horiuchi}}, \bibinfo
  {author} {\bibfnamefont {M.}~\bibnamefont {Kimura}}, \bibinfo {author}
  {\bibfnamefont {Z.}~\bibnamefont {Ren}}, \bibinfo {author} {\bibfnamefont
  {G.}~\bibnamefont {R\"opke}}, \bibinfo {author} {\bibfnamefont
  {P.}~\bibnamefont {Schuck}}, \bibinfo {author} {\bibfnamefont
  {A.}~\bibnamefont {Tohsaki}}, \bibinfo {author} {\bibfnamefont
  {C.}~\bibnamefont {Xu}}, \ and\ \bibinfo {author} {\bibfnamefont
  {T.}~\bibnamefont {Yamada}},\ }\href {\doibase 10.1103/PhysRevC.99.051303}
  {\bibfield  {journal} {\bibinfo  {journal} {Phys. Rev. C}\ }\textbf {\bibinfo
  {volume} {99}},\ \bibinfo {pages} {051303} (\bibinfo {year} {2019})},\
  \Eprint {http://arxiv.org/abs/1904.07751} {arXiv:1904.07751 [nucl-th]}
  \BibitemShut {NoStop}%
\bibitem [{\citenamefont {Li}\ \emph {et~al.}(2020{\natexlab{b}})\citenamefont
  {Li}, \citenamefont {Myo}, \citenamefont {Zhao}, \citenamefont {Toki},
  \citenamefont {Horiuchi}, \citenamefont {Xu}, \citenamefont {Liu},
  \citenamefont {Lyu},\ and\ \citenamefont {Ren}}]{Li:2020exz}%
  \BibitemOpen
  \bibfield  {author} {\bibinfo {author} {\bibfnamefont {S.}~\bibnamefont
  {Li}}, \bibinfo {author} {\bibfnamefont {T.}~\bibnamefont {Myo}}, \bibinfo
  {author} {\bibfnamefont {Q.}~\bibnamefont {Zhao}}, \bibinfo {author}
  {\bibfnamefont {H.}~\bibnamefont {Toki}}, \bibinfo {author} {\bibfnamefont
  {H.}~\bibnamefont {Horiuchi}}, \bibinfo {author} {\bibfnamefont
  {C.}~\bibnamefont {Xu}}, \bibinfo {author} {\bibfnamefont {J.}~\bibnamefont
  {Liu}}, \bibinfo {author} {\bibfnamefont {M.}~\bibnamefont {Lyu}}, \ and\
  \bibinfo {author} {\bibfnamefont {Z.}~\bibnamefont {Ren}},\ }\href {\doibase
  10.1103/PhysRevC.101.064307} {\bibfield  {journal} {\bibinfo  {journal}
  {Phys. Rev. C}\ }\textbf {\bibinfo {volume} {101}},\ \bibinfo {pages}
  {064307} (\bibinfo {year} {2020}{\natexlab{b}})},\ \Eprint
  {http://arxiv.org/abs/2005.04409} {arXiv:2005.04409 [nucl-th]} \BibitemShut
  {NoStop}%
\bibitem [{\citenamefont {Typel}(2014)}]{Typel:2014tqa}%
  \BibitemOpen
  \bibfield  {author} {\bibinfo {author} {\bibfnamefont {S.}~\bibnamefont
  {Typel}},\ }\href {\doibase 10.1103/PhysRevC.89.064321} {\bibfield  {journal}
  {\bibinfo  {journal} {Phys. Rev. C}\ }\textbf {\bibinfo {volume} {89}},\
  \bibinfo {pages} {064321} (\bibinfo {year} {2014})},\ \Eprint
  {http://arxiv.org/abs/1403.2851} {arXiv:1403.2851 [nucl-th]} \BibitemShut
  {NoStop}%
\bibitem [{\citenamefont {Adelberger}\ \emph {et~al.}(2011)\citenamefont
  {Adelberger} \emph {et~al.}}]{Adelberger:2010qa}%
  \BibitemOpen
  \bibfield  {author} {\bibinfo {author} {\bibfnamefont {E.~G.}\ \bibnamefont
  {Adelberger}} \emph {et~al.},\ }\href {\doibase 10.1103/RevModPhys.83.195}
  {\bibfield  {journal} {\bibinfo  {journal} {Rev. Mod. Phys.}\ }\textbf
  {\bibinfo {volume} {83}},\ \bibinfo {pages} {195} (\bibinfo {year} {2011})},\
  \Eprint {http://arxiv.org/abs/1004.2318} {arXiv:1004.2318 [nucl-ex]}
  \BibitemShut {NoStop}%
\bibitem [{\citenamefont {Wiescher}\ \emph {et~al.}(2010)\citenamefont
  {Wiescher}, \citenamefont {Gorres}, \citenamefont {Uberseder}, \citenamefont
  {Imbriani},\ and\ \citenamefont {Pignatari}}]{Wiescher:2010zz}%
  \BibitemOpen
  \bibfield  {author} {\bibinfo {author} {\bibfnamefont {M.}~\bibnamefont
  {Wiescher}}, \bibinfo {author} {\bibfnamefont {J.}~\bibnamefont {Gorres}},
  \bibinfo {author} {\bibfnamefont {E.}~\bibnamefont {Uberseder}}, \bibinfo
  {author} {\bibfnamefont {G.}~\bibnamefont {Imbriani}}, \ and\ \bibinfo
  {author} {\bibfnamefont {M.}~\bibnamefont {Pignatari}},\ }\href {\doibase
  10.1146/annurev.nucl.012809.104505} {\bibfield  {journal} {\bibinfo
  {journal} {Ann. Rev. Nucl. Part. Sci.}\ }\textbf {\bibinfo {volume} {60}},\
  \bibinfo {pages} {381} (\bibinfo {year} {2010})}\BibitemShut {NoStop}%
\bibitem [{\citenamefont {Coc}\ \emph {et~al.}(2012)\citenamefont {Coc},
  \citenamefont {Goriely}, \citenamefont {Xu}, \citenamefont {Saimpert},\ and\
  \citenamefont {Vangioni}}]{Coc:2011az}%
  \BibitemOpen
  \bibfield  {author} {\bibinfo {author} {\bibfnamefont {A.}~\bibnamefont
  {Coc}}, \bibinfo {author} {\bibfnamefont {S.}~\bibnamefont {Goriely}},
  \bibinfo {author} {\bibfnamefont {Y.}~\bibnamefont {Xu}}, \bibinfo {author}
  {\bibfnamefont {M.}~\bibnamefont {Saimpert}}, \ and\ \bibinfo {author}
  {\bibfnamefont {E.}~\bibnamefont {Vangioni}},\ }\href {\doibase
  10.1088/0004-637X/744/2/158} {\bibfield  {journal} {\bibinfo  {journal}
  {Astrophys. J.}\ }\textbf {\bibinfo {volume} {744}},\ \bibinfo {pages} {158}
  (\bibinfo {year} {2012})},\ \Eprint {http://arxiv.org/abs/1107.1117}
  {arXiv:1107.1117 [astro-ph.CO]} \BibitemShut {NoStop}%
\bibitem [{\citenamefont {He}\ \emph {et~al.}(2014)\citenamefont {He},
  \citenamefont {Ma}, \citenamefont {Cao}, \citenamefont {Cai},\ and\
  \citenamefont {Zhang}}]{He:2014iqa}%
  \BibitemOpen
  \bibfield  {author} {\bibinfo {author} {\bibfnamefont {W.~B.}\ \bibnamefont
  {He}}, \bibinfo {author} {\bibfnamefont {Y.~G.}\ \bibnamefont {Ma}}, \bibinfo
  {author} {\bibfnamefont {X.~G.}\ \bibnamefont {Cao}}, \bibinfo {author}
  {\bibfnamefont {X.~Z.}\ \bibnamefont {Cai}}, \ and\ \bibinfo {author}
  {\bibfnamefont {G.~Q.}\ \bibnamefont {Zhang}},\ }\href {\doibase
  10.1103/PhysRevLett.113.032506} {\bibfield  {journal} {\bibinfo  {journal}
  {Phys. Rev. Lett.}\ }\textbf {\bibinfo {volume} {113}},\ \bibinfo {pages}
  {032506} (\bibinfo {year} {2014})},\ \Eprint {http://arxiv.org/abs/1407.5414}
  {arXiv:1407.5414 [nucl-th]} \BibitemShut {NoStop}%
\bibitem [{\citenamefont {Huang}\ \emph {et~al.}(2017)\citenamefont {Huang},
  \citenamefont {Ma},\ and\ \citenamefont {He}}]{Huang:2017ysr}%
  \BibitemOpen
  \bibfield  {author} {\bibinfo {author} {\bibfnamefont {B.~S.}\ \bibnamefont
  {Huang}}, \bibinfo {author} {\bibfnamefont {Y.~G.}\ \bibnamefont {Ma}}, \
  and\ \bibinfo {author} {\bibfnamefont {W.~B.}\ \bibnamefont {He}},\ }\href
  {\doibase 10.1103/PhysRevC.95.034606} {\bibfield  {journal} {\bibinfo
  {journal} {Phys. Rev. C}\ }\textbf {\bibinfo {volume} {95}},\ \bibinfo
  {pages} {034606} (\bibinfo {year} {2017})},\ \Eprint
  {http://arxiv.org/abs/1803.07972} {arXiv:1803.07972 [nucl-th]} \BibitemShut
  {NoStop}%
\bibitem [{\citenamefont {Shi}\ and\ \citenamefont {Ma}(2021)}]{Shi:2021far}%
  \BibitemOpen
  \bibfield  {author} {\bibinfo {author} {\bibfnamefont {C.-Z.}\ \bibnamefont
  {Shi}}\ and\ \bibinfo {author} {\bibfnamefont {Y.-G.}\ \bibnamefont {Ma}},\
  }\href {\doibase 10.1007/s41365-021-00897-9} {\bibfield  {journal} {\bibinfo
  {journal} {Nucl. Sci. Tech.}\ }\textbf {\bibinfo {volume} {32}},\ \bibinfo
  {pages} {66} (\bibinfo {year} {2021})},\ \Eprint
  {http://arxiv.org/abs/2109.09938} {arXiv:2109.09938 [nucl-th]} \BibitemShut
  {NoStop}%
\bibitem [{\citenamefont {Zhang}\ \emph {et~al.}(2017)\citenamefont {Zhang},
  \citenamefont {Ma}, \citenamefont {Chen}, \citenamefont {He},\ and\
  \citenamefont {Zhong}}]{PhysRevC.95.064904}%
  \BibitemOpen
  \bibfield  {author} {\bibinfo {author} {\bibfnamefont {S.}~\bibnamefont
  {Zhang}}, \bibinfo {author} {\bibfnamefont {Y.~G.}\ \bibnamefont {Ma}},
  \bibinfo {author} {\bibfnamefont {J.~H.}\ \bibnamefont {Chen}}, \bibinfo
  {author} {\bibfnamefont {W.~B.}\ \bibnamefont {He}}, \ and\ \bibinfo {author}
  {\bibfnamefont {C.}~\bibnamefont {Zhong}},\ }\href {\doibase
  10.1103/PhysRevC.95.064904} {\bibfield  {journal} {\bibinfo  {journal} {Phys.
  Rev. C}\ }\textbf {\bibinfo {volume} {95}},\ \bibinfo {pages} {064904}
  (\bibinfo {year} {2017})}\BibitemShut {NoStop}%
\bibitem [{\citenamefont {Rybczy\ifmmode~\acute{n}\else \'{n}\fi{}ski}\ \emph
  {et~al.}(2018)\citenamefont {Rybczy\ifmmode~\acute{n}\else \'{n}\fi{}ski},
  \citenamefont {Piotrowska},\ and\ \citenamefont
  {Broniowski}}]{PhysRevC.97.034912}%
  \BibitemOpen
  \bibfield  {author} {\bibinfo {author} {\bibfnamefont {M.}~\bibnamefont
  {Rybczy\ifmmode~\acute{n}\else \'{n}\fi{}ski}}, \bibinfo {author}
  {\bibfnamefont {M.}~\bibnamefont {Piotrowska}}, \ and\ \bibinfo {author}
  {\bibfnamefont {W.}~\bibnamefont {Broniowski}},\ }\href {\doibase
  10.1103/PhysRevC.97.034912} {\bibfield  {journal} {\bibinfo  {journal} {Phys.
  Rev. C}\ }\textbf {\bibinfo {volume} {97}},\ \bibinfo {pages} {034912}
  (\bibinfo {year} {2018})}\BibitemShut {NoStop}%
\bibitem [{\citenamefont {Behera}\ \emph {et~al.}(2023)\citenamefont {Behera},
  \citenamefont {Prasad}, \citenamefont {Mallick},\ and\ \citenamefont
  {Sahoo}}]{Behera:2023nwj}%
  \BibitemOpen
  \bibfield  {author} {\bibinfo {author} {\bibfnamefont {D.}~\bibnamefont
  {Behera}}, \bibinfo {author} {\bibfnamefont {S.}~\bibnamefont {Prasad}},
  \bibinfo {author} {\bibfnamefont {N.}~\bibnamefont {Mallick}}, \ and\
  \bibinfo {author} {\bibfnamefont {R.}~\bibnamefont {Sahoo}},\ }\href
  {\doibase 10.1103/PhysRevD.108.054022} {\bibfield  {journal} {\bibinfo
  {journal} {Phys. Rev. D}\ }\textbf {\bibinfo {volume} {108}},\ \bibinfo
  {pages} {054022} (\bibinfo {year} {2023})},\ \Eprint
  {http://arxiv.org/abs/2304.10879} {arXiv:2304.10879 [hep-ph]} \BibitemShut
  {NoStop}%
\bibitem [{\citenamefont {Summerfield}\ \emph {et~al.}(2021)\citenamefont
  {Summerfield}, \citenamefont {Lu}, \citenamefont {Plumberg}, \citenamefont
  {Lee}, \citenamefont {Noronha-Hostler},\ and\ \citenamefont
  {Timmins}}]{PhysRevC.104.L041901}%
  \BibitemOpen
  \bibfield  {author} {\bibinfo {author} {\bibfnamefont {N.}~\bibnamefont
  {Summerfield}}, \bibinfo {author} {\bibfnamefont {B.-N.}\ \bibnamefont {Lu}},
  \bibinfo {author} {\bibfnamefont {C.}~\bibnamefont {Plumberg}}, \bibinfo
  {author} {\bibfnamefont {D.}~\bibnamefont {Lee}}, \bibinfo {author}
  {\bibfnamefont {J.}~\bibnamefont {Noronha-Hostler}}, \ and\ \bibinfo {author}
  {\bibfnamefont {A.}~\bibnamefont {Timmins}},\ }\href {\doibase
  10.1103/PhysRevC.104.L041901} {\bibfield  {journal} {\bibinfo  {journal}
  {Phys. Rev. C}\ }\textbf {\bibinfo {volume} {104}},\ \bibinfo {pages}
  {L041901} (\bibinfo {year} {2021})}\BibitemShut {NoStop}%
\bibitem [{\citenamefont {Tarafdar}\ \emph {et~al.}(2014)\citenamefont
  {Tarafdar}, \citenamefont {Citron},\ and\ \citenamefont
  {Milov}}]{Tarafdar:2014oua}%
  \BibitemOpen
  \bibfield  {author} {\bibinfo {author} {\bibfnamefont {S.}~\bibnamefont
  {Tarafdar}}, \bibinfo {author} {\bibfnamefont {Z.}~\bibnamefont {Citron}}, \
  and\ \bibinfo {author} {\bibfnamefont {A.}~\bibnamefont {Milov}},\ }\href
  {\doibase 10.1016/j.nima.2014.09.060} {\bibfield  {journal} {\bibinfo
  {journal} {Nucl. Instrum. Meth. A}\ }\textbf {\bibinfo {volume} {768}},\
  \bibinfo {pages} {170} (\bibinfo {year} {2014})},\ \Eprint
  {http://arxiv.org/abs/1405.4555} {arXiv:1405.4555 [nucl-ex]} \BibitemShut
  {NoStop}%
\bibitem [{\citenamefont {Volkov}(1965)}]{Volkov_1965_NP}%
  \BibitemOpen
  \bibfield  {author} {\bibinfo {author} {\bibfnamefont {A.}~\bibnamefont
  {Volkov}},\ }\href {\doibase 10.1016/0029-5582(65)90244-0} {\bibfield
  {journal} {\bibinfo  {journal} {Nuclear Physics}\ }\textbf {\bibinfo {volume}
  {74}},\ \bibinfo {pages} {33} (\bibinfo {year} {1965})}\BibitemShut {NoStop}%
\bibitem [{\citenamefont {Itagaki}\ \emph {et~al.}(2000)\citenamefont
  {Itagaki}, \citenamefont {Okabe},\ and\ \citenamefont
  {Ikeda}}]{Itagaki_2000_PRC}%
  \BibitemOpen
  \bibfield  {author} {\bibinfo {author} {\bibfnamefont {N.}~\bibnamefont
  {Itagaki}}, \bibinfo {author} {\bibfnamefont {S.}~\bibnamefont {Okabe}}, \
  and\ \bibinfo {author} {\bibfnamefont {K.}~\bibnamefont {Ikeda}},\ }\href
  {\doibase 10.1103/PhysRevC.62.034301} {\bibfield  {journal} {\bibinfo
  {journal} {Phys. Rev. C}\ }\textbf {\bibinfo {volume} {62}},\ \bibinfo
  {pages} {034301} (\bibinfo {year} {2000})}\BibitemShut {NoStop}%
\bibitem [{\citenamefont {Itagaki}\ \emph {et~al.}(1995)\citenamefont
  {Itagaki}, \citenamefont {Ohnishi},\ and\ \citenamefont
  {Kato}}]{Itagaki_1995_PTP}%
  \BibitemOpen
  \bibfield  {author} {\bibinfo {author} {\bibfnamefont {N.}~\bibnamefont
  {Itagaki}}, \bibinfo {author} {\bibfnamefont {A.}~\bibnamefont {Ohnishi}}, \
  and\ \bibinfo {author} {\bibfnamefont {K.}~\bibnamefont {Kato}},\ }\href
  {\doibase 10.1143/PTP.94.1019} {\bibfield  {journal} {\bibinfo  {journal}
  {Progress of Theoretical Physics}\ } (\bibinfo {year} {1995}),\
  10.1143/PTP.94.1019}\BibitemShut {NoStop}%
\bibitem [{\citenamefont {Ring}\ and\ \citenamefont
  {Schuck}(2004)}]{Ring_2004_Book}%
  \BibitemOpen
  \bibfield  {author} {\bibinfo {author} {\bibfnamefont {P.}~\bibnamefont
  {Ring}}\ and\ \bibinfo {author} {\bibfnamefont {P.}~\bibnamefont {Schuck}},\
  }\href@noop {} {\emph {\bibinfo {title} {The Nuclear Many-Body Problem}}}\
  (\bibinfo  {publisher} {Springer Berlin, Heidelberg},\ \bibinfo {year}
  {2004})\BibitemShut {NoStop}%
\bibitem [{\citenamefont {Chen}\ \emph {et~al.}(2010)\citenamefont {Chen},
  \citenamefont {Ko}, \citenamefont {Li},\ and\ \citenamefont
  {Xu}}]{Chen:2010qx}%
  \BibitemOpen
  \bibfield  {author} {\bibinfo {author} {\bibfnamefont {L.-W.}\ \bibnamefont
  {Chen}}, \bibinfo {author} {\bibfnamefont {C.~M.}\ \bibnamefont {Ko}},
  \bibinfo {author} {\bibfnamefont {B.-A.}\ \bibnamefont {Li}}, \ and\ \bibinfo
  {author} {\bibfnamefont {J.}~\bibnamefont {Xu}},\ }\href {\doibase
  10.1103/PhysRevC.82.024321} {\bibfield  {journal} {\bibinfo  {journal} {Phys.
  Rev. C}\ }\textbf {\bibinfo {volume} {82}},\ \bibinfo {pages} {024321}
  (\bibinfo {year} {2010})},\ \Eprint {http://arxiv.org/abs/1004.4672}
  {arXiv:1004.4672 [nucl-th]} \BibitemShut {NoStop}%
\bibitem [{\citenamefont {Angeli}(2004)}]{Angeli:2004kvy}%
  \BibitemOpen
  \bibfield  {author} {\bibinfo {author} {\bibfnamefont {I.}~\bibnamefont
  {Angeli}},\ }\href {\doibase 10.1016/j.adt.2004.04.002} {\bibfield  {journal}
  {\bibinfo  {journal} {Atom. Data Nucl. Data Tabl.}\ }\textbf {\bibinfo
  {volume} {87}},\ \bibinfo {pages} {185} (\bibinfo {year} {2004})}\BibitemShut
  {NoStop}%
\bibitem [{\citenamefont {Lin}\ \emph {et~al.}(2005)\citenamefont {Lin},
  \citenamefont {Ko}, \citenamefont {Li}, \citenamefont {Zhang},\ and\
  \citenamefont {Pal}}]{Lin:2004en}%
  \BibitemOpen
  \bibfield  {author} {\bibinfo {author} {\bibfnamefont {Z.-W.}\ \bibnamefont
  {Lin}}, \bibinfo {author} {\bibfnamefont {C.~M.}\ \bibnamefont {Ko}},
  \bibinfo {author} {\bibfnamefont {B.-A.}\ \bibnamefont {Li}}, \bibinfo
  {author} {\bibfnamefont {B.}~\bibnamefont {Zhang}}, \ and\ \bibinfo {author}
  {\bibfnamefont {S.}~\bibnamefont {Pal}},\ }\href {\doibase
  10.1103/PhysRevC.72.064901} {\bibfield  {journal} {\bibinfo  {journal} {Phys.
  Rev. C}\ }\textbf {\bibinfo {volume} {72}},\ \bibinfo {pages} {064901}
  (\bibinfo {year} {2005})},\ \Eprint {http://arxiv.org/abs/nucl-th/0411110}
  {arXiv:nucl-th/0411110} \BibitemShut {NoStop}%
\bibitem [{\citenamefont {Miller}\ \emph {et~al.}(2007)\citenamefont {Miller},
  \citenamefont {Reygers}, \citenamefont {Sanders},\ and\ \citenamefont
  {Steinberg}}]{Miller:2007ri}%
  \BibitemOpen
  \bibfield  {author} {\bibinfo {author} {\bibfnamefont {M.~L.}\ \bibnamefont
  {Miller}}, \bibinfo {author} {\bibfnamefont {K.}~\bibnamefont {Reygers}},
  \bibinfo {author} {\bibfnamefont {S.~J.}\ \bibnamefont {Sanders}}, \ and\
  \bibinfo {author} {\bibfnamefont {P.}~\bibnamefont {Steinberg}},\ }\href
  {\doibase 10.1146/annurev.nucl.57.090506.123020} {\bibfield  {journal}
  {\bibinfo  {journal} {Ann. Rev. Nucl. Part. Sci.}\ }\textbf {\bibinfo
  {volume} {57}},\ \bibinfo {pages} {205} (\bibinfo {year} {2007})},\ \Eprint
  {http://arxiv.org/abs/nucl-ex/0701025} {arXiv:nucl-ex/0701025} \BibitemShut
  {NoStop}%
\bibitem [{\citenamefont {Kharzeev}\ and\ \citenamefont
  {Nardi}(2001)}]{Kharzeev:2000ph}%
  \BibitemOpen
  \bibfield  {author} {\bibinfo {author} {\bibfnamefont {D.}~\bibnamefont
  {Kharzeev}}\ and\ \bibinfo {author} {\bibfnamefont {M.}~\bibnamefont
  {Nardi}},\ }\href {\doibase 10.1016/S0370-2693(01)00457-9} {\bibfield
  {journal} {\bibinfo  {journal} {Phys. Lett. B}\ }\textbf {\bibinfo {volume}
  {507}},\ \bibinfo {pages} {121} (\bibinfo {year} {2001})},\ \Eprint
  {http://arxiv.org/abs/nucl-th/0012025} {arXiv:nucl-th/0012025} \BibitemShut
  {NoStop}%
\bibitem [{\citenamefont {Loizides}\ \emph {et~al.}(2018)\citenamefont
  {Loizides}, \citenamefont {Kamin},\ and\ \citenamefont
  {d'Enterria}}]{Loizides:2017ack}%
  \BibitemOpen
  \bibfield  {author} {\bibinfo {author} {\bibfnamefont {C.}~\bibnamefont
  {Loizides}}, \bibinfo {author} {\bibfnamefont {J.}~\bibnamefont {Kamin}}, \
  and\ \bibinfo {author} {\bibfnamefont {D.}~\bibnamefont {d'Enterria}},\
  }\href {\doibase 10.1103/PhysRevC.97.054910} {\bibfield  {journal} {\bibinfo
  {journal} {Phys. Rev. C}\ }\textbf {\bibinfo {volume} {97}},\ \bibinfo
  {pages} {054910} (\bibinfo {year} {2018})},\ \bibinfo {note} {[Erratum:
  Phys.Rev.C 99, 019901 (2019)]},\ \Eprint {http://arxiv.org/abs/1710.07098}
  {arXiv:1710.07098 [nucl-ex]} \BibitemShut {NoStop}%
\bibitem [{\citenamefont {Li}\ \emph {et~al.}(1997)\citenamefont {Li},
  \citenamefont {Ko},\ and\ \citenamefont {Ren}}]{Li:1997rc}%
  \BibitemOpen
  \bibfield  {author} {\bibinfo {author} {\bibfnamefont {B.-A.}\ \bibnamefont
  {Li}}, \bibinfo {author} {\bibfnamefont {C.~M.}\ \bibnamefont {Ko}}, \ and\
  \bibinfo {author} {\bibfnamefont {Z.-z.}\ \bibnamefont {Ren}},\ }\href
  {\doibase 10.1103/PhysRevLett.78.1644} {\bibfield  {journal} {\bibinfo
  {journal} {Phys. Rev. Lett.}\ }\textbf {\bibinfo {volume} {78}},\ \bibinfo
  {pages} {1644} (\bibinfo {year} {1997})},\ \Eprint
  {http://arxiv.org/abs/nucl-th/9701048} {arXiv:nucl-th/9701048} \BibitemShut
  {NoStop}%
\bibitem [{\citenamefont {Chen}\ \emph {et~al.}(2003)\citenamefont {Chen},
  \citenamefont {Ko},\ and\ \citenamefont {Li}}]{Chen:2003ava}%
  \BibitemOpen
  \bibfield  {author} {\bibinfo {author} {\bibfnamefont {L.-W.}\ \bibnamefont
  {Chen}}, \bibinfo {author} {\bibfnamefont {C.~M.}\ \bibnamefont {Ko}}, \ and\
  \bibinfo {author} {\bibfnamefont {B.-A.}\ \bibnamefont {Li}},\ }\href
  {\doibase 10.1016/j.nuclphysa.2003.09.010} {\bibfield  {journal} {\bibinfo
  {journal} {Nucl. Phys. A}\ }\textbf {\bibinfo {volume} {729}},\ \bibinfo
  {pages} {809} (\bibinfo {year} {2003})},\ \Eprint
  {http://arxiv.org/abs/nucl-th/0306032} {arXiv:nucl-th/0306032} \BibitemShut
  {NoStop}%
\bibitem [{\citenamefont {Sun}\ and\ \citenamefont {Chen}(2017)}]{Sun:2017ooe}%
  \BibitemOpen
  \bibfield  {author} {\bibinfo {author} {\bibfnamefont {K.-J.}\ \bibnamefont
  {Sun}}\ and\ \bibinfo {author} {\bibfnamefont {L.-W.}\ \bibnamefont {Chen}},\
  }\href {\doibase 10.1103/PhysRevC.95.044905} {\bibfield  {journal} {\bibinfo
  {journal} {Phys. Rev. C}\ }\textbf {\bibinfo {volume} {95}},\ \bibinfo
  {pages} {044905} (\bibinfo {year} {2017})},\ \Eprint
  {http://arxiv.org/abs/1701.01935} {arXiv:1701.01935 [nucl-th]} \BibitemShut
  {NoStop}%
\bibitem [{\citenamefont {Charity}\ \emph {et~al.}(1988)\citenamefont {Charity}
  \emph {et~al.}}]{Charity:1988zz}%
  \BibitemOpen
  \bibfield  {author} {\bibinfo {author} {\bibfnamefont {R.~J.}\ \bibnamefont
  {Charity}} \emph {et~al.},\ }\href {\doibase 10.1016/0375-9474(88)90542-8}
  {\bibfield  {journal} {\bibinfo  {journal} {Nucl. Phys. A}\ }\textbf
  {\bibinfo {volume} {483}},\ \bibinfo {pages} {371} (\bibinfo {year}
  {1988})}\BibitemShut {NoStop}%
\bibitem [{\citenamefont {Charity}(2010)}]{Charity:2010wk}%
  \BibitemOpen
  \bibfield  {author} {\bibinfo {author} {\bibfnamefont {R.~J.}\ \bibnamefont
  {Charity}},\ }\href {\doibase 10.1103/PhysRevC.82.014610} {\bibfield
  {journal} {\bibinfo  {journal} {Phys. Rev. C}\ }\textbf {\bibinfo {volume}
  {82}},\ \bibinfo {pages} {014610} (\bibinfo {year} {2010})},\ \Eprint
  {http://arxiv.org/abs/1006.5018} {arXiv:1006.5018 [nucl-th]} \BibitemShut
  {NoStop}%
\bibitem [{\citenamefont {Wong}(1982)}]{Wong:1982zzb}%
  \BibitemOpen
  \bibfield  {author} {\bibinfo {author} {\bibfnamefont {C.-Y.}\ \bibnamefont
  {Wong}},\ }\href {\doibase 10.1103/PhysRevC.25.1460} {\bibfield  {journal}
  {\bibinfo  {journal} {Phys. Rev. C}\ }\textbf {\bibinfo {volume} {25}},\
  \bibinfo {pages} {1460} (\bibinfo {year} {1982})}\BibitemShut {NoStop}%
\bibitem [{\citenamefont {Bertsch}\ and\ \citenamefont
  {Das~Gupta}(1988)}]{Bertsch:1988ik}%
  \BibitemOpen
  \bibfield  {author} {\bibinfo {author} {\bibfnamefont {G.~F.}\ \bibnamefont
  {Bertsch}}\ and\ \bibinfo {author} {\bibfnamefont {S.}~\bibnamefont
  {Das~Gupta}},\ }\href {\doibase 10.1016/0370-1573(88)90170-6} {\bibfield
  {journal} {\bibinfo  {journal} {Phys. Rept.}\ }\textbf {\bibinfo {volume}
  {160}},\ \bibinfo {pages} {189} (\bibinfo {year} {1988})}\BibitemShut
  {NoStop}%
\bibitem [{\citenamefont {Wang}\ \emph {et~al.}(2021)\citenamefont {Wang},
  \citenamefont {Huang}, \citenamefont {Kondev}, \citenamefont {Audi},\ and\
  \citenamefont {Naimi}}]{Wang:2021xhn}%
  \BibitemOpen
  \bibfield  {author} {\bibinfo {author} {\bibfnamefont {M.}~\bibnamefont
  {Wang}}, \bibinfo {author} {\bibfnamefont {W.~J.}\ \bibnamefont {Huang}},
  \bibinfo {author} {\bibfnamefont {F.~G.}\ \bibnamefont {Kondev}}, \bibinfo
  {author} {\bibfnamefont {G.}~\bibnamefont {Audi}}, \ and\ \bibinfo {author}
  {\bibfnamefont {S.}~\bibnamefont {Naimi}},\ }\href {\doibase
  10.1088/1674-1137/abddaf} {\bibfield  {journal} {\bibinfo  {journal} {Chin.
  Phys. C}\ }\textbf {\bibinfo {volume} {45}},\ \bibinfo {pages} {030003}
  (\bibinfo {year} {2021})}\BibitemShut {NoStop}%
\bibitem [{\citenamefont {Wang}\ \emph {et~al.}(2014)\citenamefont {Wang},
  \citenamefont {Liu}, \citenamefont {Wu},\ and\ \citenamefont
  {Meng}}]{Wang:2014qqa}%
  \BibitemOpen
  \bibfield  {author} {\bibinfo {author} {\bibfnamefont {N.}~\bibnamefont
  {Wang}}, \bibinfo {author} {\bibfnamefont {M.}~\bibnamefont {Liu}}, \bibinfo
  {author} {\bibfnamefont {X.}~\bibnamefont {Wu}}, \ and\ \bibinfo {author}
  {\bibfnamefont {J.}~\bibnamefont {Meng}},\ }\href {\doibase
  10.1016/j.physletb.2014.05.049} {\bibfield  {journal} {\bibinfo  {journal}
  {Phys. Lett. B}\ }\textbf {\bibinfo {volume} {734}},\ \bibinfo {pages} {215}
  (\bibinfo {year} {2014})},\ \Eprint {http://arxiv.org/abs/1405.2616}
  {arXiv:1405.2616 [nucl-th]} \BibitemShut {NoStop}%
\end{thebibliography}%
\end{document}